\documentclass[article]{IEEEtran}

\ifCLASSINFOpdf
	\usepackage[pdftex]{graphicx}
    \graphicspath{{Figures/}}
    \DeclareGraphicsExtensions{.pdf}
\else
\fi
\usepackage{graphicx}
\ifCLASSOPTIONcompsoc
  \usepackage[caption=false,font=normalsize,labelfont=sf,textfont=sf]{subfig}
\else
  \usepackage[caption=false,font=footnotesize]{subfig}
\fi

\usepackage{amsfonts}
\usepackage{amsmath}
\makeatletter
\def\Let@{\def\\{\notag\math@cr}}
\makeatother
\usepackage{colortbl}
\usepackage{cite}
\usepackage{booktabs}
\usepackage{multirow}
\usepackage{algorithm} 
\usepackage{algorithmicx}
\usepackage{algpseudocode}
\usepackage[shortlabels]{enumitem}
\usepackage{accents}
\usepackage{color}

\begin{document}
	
\title{Beyond Phasors: Continuous-Spectrum Modeling\\of Power Systems using the Hilbert Transform}

\author{Asja~Dervi{\v s}kadi{\'c},  Guglielmo~Frigo,  Mario~Paolone
}

\maketitle

\begin{abstract}
Modern power systems are at risk of largely reducing the inertia of generation assets and prone to experience extreme dynamics. 
The consequence is that, during electromechanical transients triggered by large contingencies, transmission of electrical power may take place in a broad spectrum well beyond the single fundamental component. 
Traditional modeling approaches rely on the phasor representation derived from the Fourier Transform (FT) of the signal under analysis. 
During large transients, though, FT-based analysis may fail to accurately identify the fundamental component parameters, in terms of amplitude, frequency and phase. 
In this paper, we propose an alternative approach relying on the Hilbert Transform (HT), that, in view of the possibility to identify the whole spectrum, enables the tracking of signal dynamics. 
We compare FT- and HT-based approaches during representative operating conditions, i.e., amplitude modulations, frequency ramps and step changes, in synthetic and real-world datasets. We further validate the approaches using a contingency analysis on the IEEE 39-bus.
\end{abstract}


\IEEEpeerreviewmaketitle

\section{Introduction}
\label{sec:intro}

Power systems are rapidly evolving towards low-inertia networks and system operators are facing new challenges to operate their grids safely \cite{Milano-etAl2018-PSCC,Winter-etAl2015-PEM,Tielens-etAl2016,Kroposki-etAl2017-PEM,Gross-etAl2017-IFAC}. 
Specifically, they are facing a dramatic increase in renewable energy sources and inverter-connected devices that, as such, do not provide any inertia to filter dynamics originated by power system disturbances  \cite{Sevilla-etAl2017-ISGT,Pulgar-etAl2018,Nguyen-etAl2018,Tuffner-etAl2019-TSG}. 
In the so-formed inverter-dominated power grids, phenomena that used to be exceptional in traditional networks, such as frequency modulations, rapid (i.e., sub-second) frequency variations or sudden amplitude steps, are more likely to be experienced, and have been identified and documented by systems' operators \cite{AEMO,SIL}. 

\begin{figure}
	\centering
	\includegraphics[width=0.87\columnwidth]{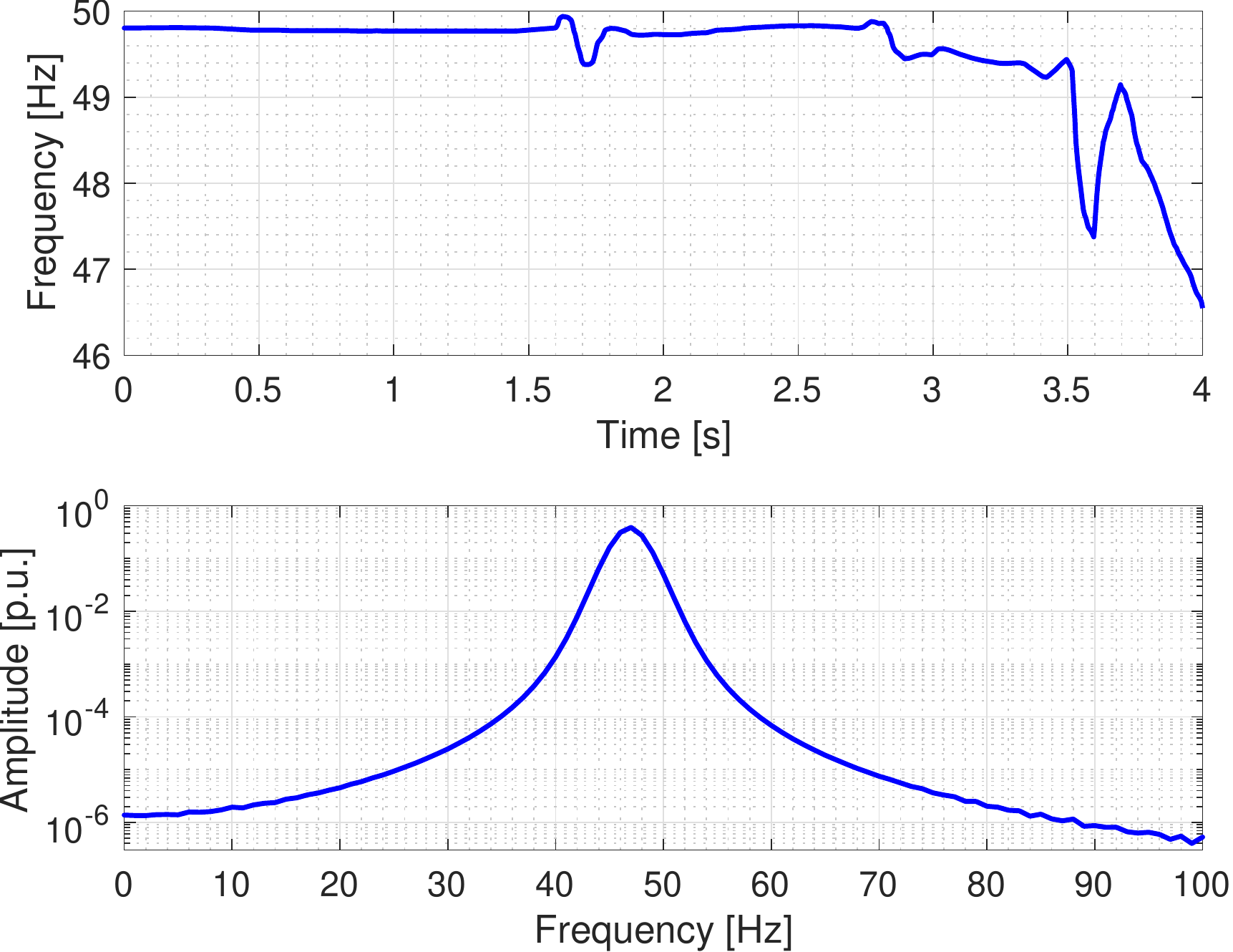}
	\caption{Frequency estimated by PMUs during the Australian blackout on September 28, 2016 \cite{AEMO} (a) and frequency spectrum of a single-tone signal characterized by ROCOF of -6.25 Hz/s computed using the FT. }
	\label{fig:intro}
\end{figure}

A clear example happened in September 2016 when the South Australian system faced a severe blackout because a strong windstorm hit the region while half of its power consumption was fed by wind generation \cite{AEMO}. 
Figure \ref{fig:intro}(a) shows the frequency as recorded by Phasor Measurement Units (PMUs) installed in that area. The frequency experienced a large drop of almost 4 Hz in about 0.7 seconds, with an estimated Rate-of-Change-of-Frequency (ROCOF) of roughly -6.25 Hz/s. 
From a theoretical perspective, it is worth investigating whether these operating conditions can be thoroughly described using traditional power system analysis tools that refer to the concept of phasor and assume signals to be in steady-state and characterized by narrow-banded spectra \cite{Phadke-2017}. 
Typically, phasor parameters are determined using the Fourier Transform (FT): the frequency, amplitude and phase of the phasor associated to the fundamental component can be inferred from the FT coefficients laying around the nominal system frequency \cite{DeLaRee-etAl2010}. 
However, during dynamic events, the FT-based representation may not allow for unequivocally identifying the parameters of the fundamental component \cite{PhadkeTransient2008}. 
The same holds even when applying frequency tracking techniques, such as the Short-Time FT (STFT)
\cite{Macii-etAl2018-AMPS}.
In this respect, Fig. \ref{fig:intro}(b) represents the spectrum of a single-tone signal at 50 Hz, whose frequency is decaying with a ROCOF of -6.25 Hz/s, thus replicating the same condition that occurred in South Australia before the blackout.
The figure shows that only 32\% of energy is transmitted in the range between 48 and 52 Hz, and that the rest is spread well beyond the fundamental tone.

The traditional approach towards power system analysis, on the one hand, has driven the development of network modeling techniques based on the concept of phasor. In this sense, the phasor extraction operation can be seen as a signal compression, driven by the necessity of providing a concise and exhaustive description of the information contained in the time-domain signals. 
On the other hand, the phasor concept has motivated also the implementation of system-awareness applications (e.g., PMUs and state estimators).
However, the capability of accurately describing the power system behaviour is strongly related to the paradigm employed to formulate its governing physical laws. 
In this regard, power system operators are nowadays experiencing difficulties in interpreting phasor data estimated during large transients and are concerned about potentially wrong control actions relaying those estimates. 
The emerging trend is to build situational awareness systems that leverage on the raw time-domain data. 
For instance, many PMU vendors are considering the idea of transferring the so-called point-on-wave data along with the PMU data stream. 
Those data could be collected at the data concentration point and then further analyzed by highly powerful computation tools. 
However this solution, besides being extremely bandwidth-demanding for the underlying telecommunication infrastructure, is simply postponing the problem of successfully compressing the information related to the considered power system event. 
In this context, we are investigating mathematical transformations that enable us to go beyond the concept of phasor based on the FT: we envision a treatise in which the signal dynamics are preserved unaltered and thus can be suitably estimated, potentially leading to novel approaches for the operation and control of power networks. 

In this regard, the Hilbert Transform (HT) might represent a promising and effective solution. Given the acquired waveform, the HT produces a complex-valued signal, also known as \textit{analytic signal}, whose spectrum consists only of real components \cite{Vetterli-2014,hahn1996hilbert}. Moreover, the HT may enable us to analytically formulate power system dynamics by means of dynamic phasors \cite{Boashash1992,Nowomiejski1981,See-etAl2012}. However, the HT can go beyond this already discussed representation in view of the possibility of defining more appropriate functional basis. 


In this paper, we discuss the advantages of an HT-based representation of power system transients.
First, we rigorously demonstrate the energy conservation property in the HT-domain. 
Then, by defining equivalent expressions of signals representing typical power system dynamics, we derive their representation in both FT- and HT-domain. Specifically, we consider (i) amplitude modulation, (ii) frequency ramp and (iii) amplitude step as representative transient operating conditions. 
The comparison of the obtained spectra enables us to highlight the limitation of FT-based analysis in non-stationary conditions and, conversely, the HT capability of accurately tracking the evolution of instantaneous power flows.
Finally, we validate these purely theoretical findings analyzing the results of dedicated time-domain simulations by means of the FT and HT. 
In particular, we use a simple 2-bus system to reproduce the three above defined transients as well as real-world datasets. We further use the IEEE 39-bus system to include a contingency analysis on a  large scale power grid.


The paper is structured as follows. 
Section \ref{sec:prob_stat} contains the problem statement and the underlying hypothesis of the proposed HT-based approach. 
Section \ref{sec:theory} provides the theoretical foundations. 
Section \ref{sec:model} describes the validation method. 
Section \ref{sec:results} assesses the proposed approach performance in representative operative conditions.
Section \ref{sec:concl} concludes the paper discussing possible implications of the proposed HT-based analysis. 

\section{Problem Statement}
\label{sec:prob_stat}

Modern power grids are characterized by voltage and current signals whose spectral content covers a bandwidth much wider than the narrow fundamental component. During dynamic events, traditional analysis tools fail in accurately identifying the signal parameters, as the FT-based decomposition provides an incomplete and approximated representation of the signal information content.
Indeed, the FT-based analysis relies on a signal model that is stationary and periodic. During dynamic events, these assumptions cannot be verified and this modeling discrepancy affects the accuracy of the FT-based representation.
The modification of the functional basis employed to model the signals coupled with the electrical circuit laws, is the only rigorous approach that can lead to a proper modeling of power system dynamics. 

Based on these consideration, we propose a novel treatise relying on the HT, that does not limit the analysis to the fundamental component bandwidth but enables us to recover the whole spectral content of the signals under analysis.
The proposed approach opens the floor for even more relevant analysis tools accounting for the bulk grid behaviour over the entire spectrum, and thus reliable not only in steady-state but also in dynamic and transient conditions.



\section{Representation of Power System Transients\\in Hilbert Transform Domain}
\label{sec:theory}

This section aims at providing the theoretical basis of our investigation. First, we lay the foundations by demonstrating the energy conservation property in the HT-domain and, therefore, proving the applicability of the HT for studying power system dynamics. 
Then, we formulate typical power system transients in the FT- and HT-domain, showing the potential of relying on an HT-based approach.

\subsection{Theoretical Foundations and Energy Conservation}
\label{sec:theory1}
For the sake of nomenclature, we recall some definitions regarding the HT. 
Given a generic time-varying real-valued signal $x(t)$, its HT $ \mathcal{H}(\cdot) $ is defined as \cite{Vetterli-2014,hahn1996hilbert}:
\begin{equation}
\tilde{x}(t) = \mathcal{H}[x(t)] = \frac{1}{\pi} \int_{-\infty}^{+\infty} \frac{x(\tau)}{t-\tau} d\tau
\label{eq:hilb}
\end{equation} 
Such transform has the property of introducing a phase shift of $-\pi/2$ at each positive frequency and $+\pi/2$ at each negative frequency. 
The combination of the real signal $x(t)$ and its HT  $\tilde{x}(t)$ form the so-called \textit{analytic signal} $ \hat{x}(t) $: 
\begin{equation}
\hat{x}(t) = x(t) + j\cdot \mathcal{H}[x(t)] = x(t) + j\cdot\tilde{x}(t)
\label{eq:ana}
\end{equation}

Let us consider two signals representing a generic voltage $v(t)$ and current $i(t)$ of a power system. The instantaneous power is computed as the product among the two waveforms: 
\begin{equation}
p(t) = v(t) \cdot i(t)
\end{equation} 
By applying \eqref{eq:ana}, the analytic signals associated to the voltage, current and instantaneous power can be expressed as: 
\begin{eqnarray}
\hat{v}(t) = v(t) + j\cdot \mathcal{H}[v(t)] = v(t) + j\cdot\tilde{v}(t)\\
\hat{i}(t) = i(t) + j\cdot \mathcal{H}[i(t)] = i(t) + j\cdot\tilde{i}(t)\\
\hat{p}(t) = p(t) + j\cdot \mathcal{H}[p(t)] = p(t) + j\cdot\tilde{p}(t)
\end{eqnarray}


Furthermore, if we compute the product between the analytic signals of voltage and current, we get:
\begin{align}
\label{eq:p1}
\hat{p}'(t) &= \hat{v}(t) \cdot \hat{i}(t) = \nonumber \\
&=[v(t) + j\cdot\tilde{v}(t) ] \cdot [i(t) + j\cdot\tilde{i}(t) ] = \nonumber \\
&= v(t) i(t) - \tilde{v}(t) \tilde{i}(t) + j \cdot [\tilde{v}(t) i(t) + v(t) \tilde{i}(t) ]
\end{align}
In a similar way, the product between the analytic signals of voltage and current complex conjugate yields: 
\begin{align}
\label{eq:p2}
\hat{p}''(t) &= \hat{v}(t) * \hat{i}(t) = \nonumber \\
&= [v(t) + j\cdot\tilde{v}(t) ] \cdot [i(t) - j\cdot\tilde{i}(t) ] = \nonumber \\
&= v(t) i(t) + \tilde{v}(t) \tilde{i}(t) + j \cdot [\tilde{v}(t) i(t) - v(t) \tilde{i}(t) ]
\end{align}
It is interesting to compute the sum of \eqref{eq:p1} and \eqref{eq:p2}:  
\begin{align}
\label{eq:p3}
\hat{p}'''(t) =& \hat{p}'(t) + \hat{p}''(t) = \\
=& 2 v(t) \cdot  i(t) + 2 j \cdot \tilde{v}(t) \cdot  i(t)
\end{align}
whose real part is proportional to the instantaneous power:
\begin{equation}
\label{eq:pht}
real(\hat{p}'''(t)) = 2p(t)
\end{equation}

Without loss of generality, voltage and current signals can be modeled as sinusoids pulsating at the same frequency $f$:
\begin{eqnarray}
v(t) = V cos(2\pi f t + \varphi_v)\\
i(t)  = I cos(2\pi f t + \varphi_i)
\end{eqnarray}
The instantaneous power is given by the waveforms' product:
\begin{equation}
\label{eq:p}
p(t) = v(t) \cdot i(t) = \frac{1}{2} VI [ cos(\varphi_v - \varphi_i) + cos(4\pi f t + \varphi_v + \varphi_i) ]
\end{equation} 
As expected, the resulting instantaneous power pulsates at a frequency that is double the power system frequency. 
The voltage and current HTs are obtained by shifting the spectra by $\pm\pi/2$, resulting in the following analytic signals:  
\begin{equation}
\hat{v}(t) = v(t) + j\cdot\tilde{v}(t) = V cos(2\pi f t + \varphi_v) + j \cdot V sin(2\pi f t + \varphi_v)
\end{equation}
\begin{equation}
\hat{i}(t)  = i(t) + j\cdot\tilde{i}(t) = I cos(2\pi f t + \varphi_i) + j \cdot I sin(2\pi f t + \varphi_i)
\end{equation}

If we compute the HT of the instantaneous power, we get:
\begin{equation}
\tilde{p}(t) = \frac{1}{2} VI sin(4\pi f t + \varphi_v + \varphi_i)
\end{equation}
However, by comparing this result with \eqref{eq:p}, we notice that the information regarding the constant offset is lost. That is to say that the HT of the instantaneous power computed by simply using its definition is useless with respect to \eqref{eq:p}. Similarly, if we compute the analytic signal of $p(t)$, we get: 
\begin{align}
\hat{p}(t) =& p(t) + j\cdot\tilde{p}(t) = \\ 
=&\frac{1}{2} VI[ cos(\varphi_v - \varphi_i) + cos(4\pi f t + \varphi_v + \varphi_i) + \\ 
&+j \cdot sin(4\pi f t + \varphi_v + \varphi_i)] 
\end{align}
that, compared to \eqref{eq:p}, maintains the constant offset information, but introduces an imaginary complex term. 

As in \eqref{eq:p1} and \eqref{eq:p2}, $\hat{p}'$ and $\hat{p}''$ can be computed as: 
\begin{equation}
\hat{p}'(t) = VI \cdot [cos(4\pi f t + \varphi_v + \varphi_i) + j \cdot sin(4\pi f t + \varphi_v + \varphi_i)]
\end{equation}
\begin{equation}
\hat{p}''(t) = VI \cdot [cos(\varphi_v - \varphi_i) + j \cdot sin(\varphi_v - \varphi_i) ]
\end{equation}
Finally, as in \eqref{eq:p3}, the sum of the obtained quantities yields to: 
\begin{align}
\hat{p}'''(t) = VI \cdot  [ cos(\varphi_v - \varphi_i) + cos(4\pi f t + \varphi_v + \varphi_i) ] + \\
j \cdot VI \cdot  [ sin(\varphi_v - \varphi_i) + sin(4\pi f t + \varphi_v + \varphi_i) ]
\end{align}
whose real part corresponds to twice the instantaneous power in \eqref{eq:p}, exactly as in \eqref{eq:pht},
demonstrating that $\hat{p}'''$ can be used to derive the instantaneous power of an electrical grid.

\subsection{Application to Realistic Scenarios}
\label{sec:theory2}
The HT enables us to analyze realistic signals typical of power system transients, where traditional FT-based techniques fail to give an appropriate phasor representation.  
For instance, let us start with the case of a generic power system signal whose amplitude is modulated by a cosine whose frequency is much slower than the fundamental one (i.e. the power system frequency). To fix ideas, this signal may represent a nodal voltage. This phenomenon typically appears during inter-area oscillations between large system regions. Formally, an amplitude modulation can be modeled as \cite{Frigo-etAl2017}: 
\begin{equation}
x(t) = A_0(1+k_a cos(2 \pi f_a t))\cdot cos(2 \pi f_0 t + \varphi_0) 
\label{eq:am}
\end{equation}
being $f_0$, $A_0$ and $\varphi_0$ the fundamental tone frequency, amplitude and initial phase, respectively, $f_a$ the modulation frequency ($f_a \ll f_0$) and $k_a$ the modulation factor. 

In this case, the signal FT can be expressed as: 
\begin{align}
\mathcal{F}[x(t)] =	& A_0/2 \cdot  \big [ \delta^-(0)e^{j\varphi_0} + \delta^+(0)e^{-j\varphi_0} + \\ 
& + k_a/2 \cdot [ \delta^-(-f_a)e^{j\varphi_0} + \delta^+(f_a)e^{-j\varphi_0} + \\
& + \delta^-(f_a)e^{j\varphi_0} + \delta^+(-f_a)e^{-j\varphi_0}] \big ] 
\end{align}
where the Dirac function $ \delta $ has been formulated as:
\begin{equation}
\delta^{\pm}(f_k) = \delta(f \pm f_0 + f_k)
\end{equation}

The spectrum is characterized by three pairs of bins in the positive and negative frequency domain: one at frequency $f_0$ and two centered around it at $f_0 \pm f_a$. In such a scenario, an FT-based signal analyzer (tailored to investigate signals around the rated power system frequency) could fail to provide a correct spectrum interpretation because of the interference generated by the modulating tones. 
Conversely, the HT-based approach enables us to derive a single component analytic signal:
\begin{equation}
\hat{x}(t) = A_0 (1 + k_a cos(2 \pi f_a t)) \cdot e^{j(2 \pi f_0 t + \varphi_0)}
\end{equation}
i.e. a phasor rotating at the fundamental frequency $f_0$, while its amplitude is pulsating at the modulation frequency $f_a$. 
Indeed, the HT of the product of two signals with non-overlapping spectra is equal to the product of the low-frequency term by the HT of the high-frequency term \cite{Vetterli-2014}.

The second case under investigation consists in a power system whose frequency is decaying with a descending ramp trend, as typical of the stages anticipating a severe system collapse. A frequency ramp can be modeled as \cite{Frigo-etAl2017}: 
\begin{equation}
x(t) = A_0 \cdot cos(2 \pi f_0 t + \varphi_0+R\pi t^2) 
\label{eq:fr}
\end{equation}
being $R$ the ramp rate. 
The signal FT can be formulated as: 
\begin{align}
\mathcal{F}[x(t)] = \frac{jRe^{-j\frac{(\omega+\omega_0)^2}{4 \pi R}+j\frac{\omega^2+\omega_0^2}{2 \pi R} -j \varphi_0}+ \sqrt{R^2}e^{-j\frac{(\omega+\omega_0)^2}{4 \pi R}+j \varphi_0}}{2/A_0 \cdot \sqrt{2\pi}\sqrt{-jR}\sqrt{R^2}}
\end{align}
where the notation $ \omega $ indicates the phasor angular velocity or pulsation. In this context, it should be noticed how difficult it is to distinguish the fundamental tone from the spurious contributions introduced by the time varying frequency. 

By contrast, the HT provides the following analytic signal:
\begin{equation}
\hat{x}(t) = A_0 \cdot e^{j(2 \pi f_0 t + \varphi_0 +R\pi t^2)}	
\end{equation}
that can be regarded as a dynamic phasor, characterized by constant amplitude and rotating with a ramping frequency.   

The final case of our investigation refers to a situation where FT-based methods provide largely discrepant results, i.e., a signal whose amplitude experiences a step, modeled as \cite{Frigo-etAl2017}:
\begin{equation}
x(t) = A_0 ( 1+k_s h(t) )\cdot  cos(2 \pi f_0 t + \varphi_0) 
\label{eq:as}
\end{equation}
being $k_s$ the step factor and $h(t)$ the Heaviside function, that is null for $t<0$ and 1 for $t \geq 0$. The signal FT is:
\begin{align}
\mathcal{F}[x(t)] = & A_0/2 \cdot [ \delta^-(0)e^{j\varphi_0} + \delta^+(0)e^{-j\varphi_0} + \\
& +  \frac{k_s(1-h(t))}{j 2\pi (f-f_0)} e^{j \varphi_0} + \frac{k_s(1-h(t))}{j 2\pi (f+f_0)}e^{-j \varphi_0} ]
\end{align}

This formulation still maintains the information regarding the fundamental tone, but contains also two hyperbolic terms whose contributions are spread over the whole frequency spectrum. In such a scenario, an FT-based approach fails in providing an appropriate reconstruction of the signal under investigation due to the scattering of the spectrum bins that largely bias the information related to the fundamental tone. 

The HT, instead, provides an analytic signal formulation:
\begin{equation}
\hat{x}(t) = A_0 (1 + k_s h(t) ) \cdot e^{j(2 \pi f_0 t + \varphi_0)}
\end{equation}
where the amplitude step information is entirely preserved.

\section{Validation Models and Algorithm}
\label{sec:model}
In this section, we provide the modeling and algorithmic details for the numerical validation of the proposed HT-based analysis. To this end, we use two models and two different simulation environments to generate datasets characterized by significant power system dynamics. 
First, we adopt a simple 2-bus model using the EMTP-RV simulation environment \cite{Mahseredjian-etAl2007,Mahseredjian2008}, in order to replicate the waveforms described in Section \ref{sec:theory}.
Second, we use the IEEE 39-bus model \cite{IEEE39bus,github39bus} implemented within the OPAL-RT environment \cite{Opal-RT}, in order to emulate the operating conditions of large-scale networks. 
We further present an algorithm that enables us to compare the performance of FT- and HT-based approaches.

\subsection{Simple 2-bus Model}
\begin{figure}
	\centering
	\includegraphics[width=0.87\columnwidth]{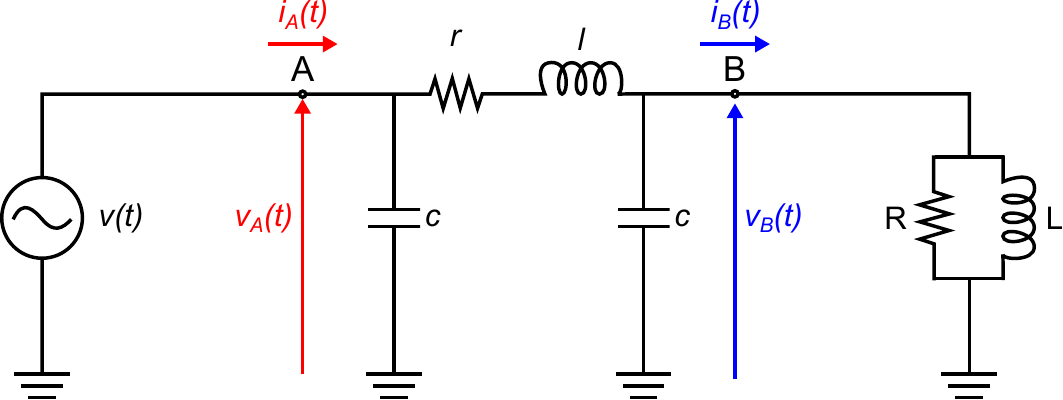}
	\caption{Block diagram of the adopted EMTP-RV simulation model. The generator is characterized by nominal voltage of 380 kV, and is connected to a 100 MW load through a 100 km ACSR overhead line.} 
	\label{fig:model}
\end{figure}

We perform off-line simulations within the EMTP-RV simulation environment \cite{Mahseredjian-etAl2007,Mahseredjian2008}.
Without loss of generality, in order to use a repeatable example, we refer to the simple model in Fig. \ref{fig:model}, 
that consists of a time-varying 3-phase voltage generator powering a generic load through a transmission line, characterized by steady-state rated voltage of 380 kV at 50 Hz.

The voltage source has been modeled by means of a look-up-table providing point-on-wave data representing a generic dynamic voltage supplying the line.
We consider a 100 km aluminum conductors steel reinforced (ACSR) overhead line characterized by commonly adopted parameters: diameter 31.5 mm, $r$ = 0.02 10$^{-3}$ $ \Omega $/m, $x$ = 0.268 10$^{-3}$ $\Omega$/m, $c$ = 13.7 10$^{-10}$ F/m. We model the transmission line using a lumped constants $\pi$ equivalent, and the load by means of an R-L parallel equivalent, where R and L have been tuned in order to get a total load of 100 MW and 0.9 power factor in case of a purely sinusoidal voltage at 50 Hz.

\subsection{The IEEE 39-bus Model}
\begin{figure}
	\centering
	\includegraphics[width=0.87\columnwidth]{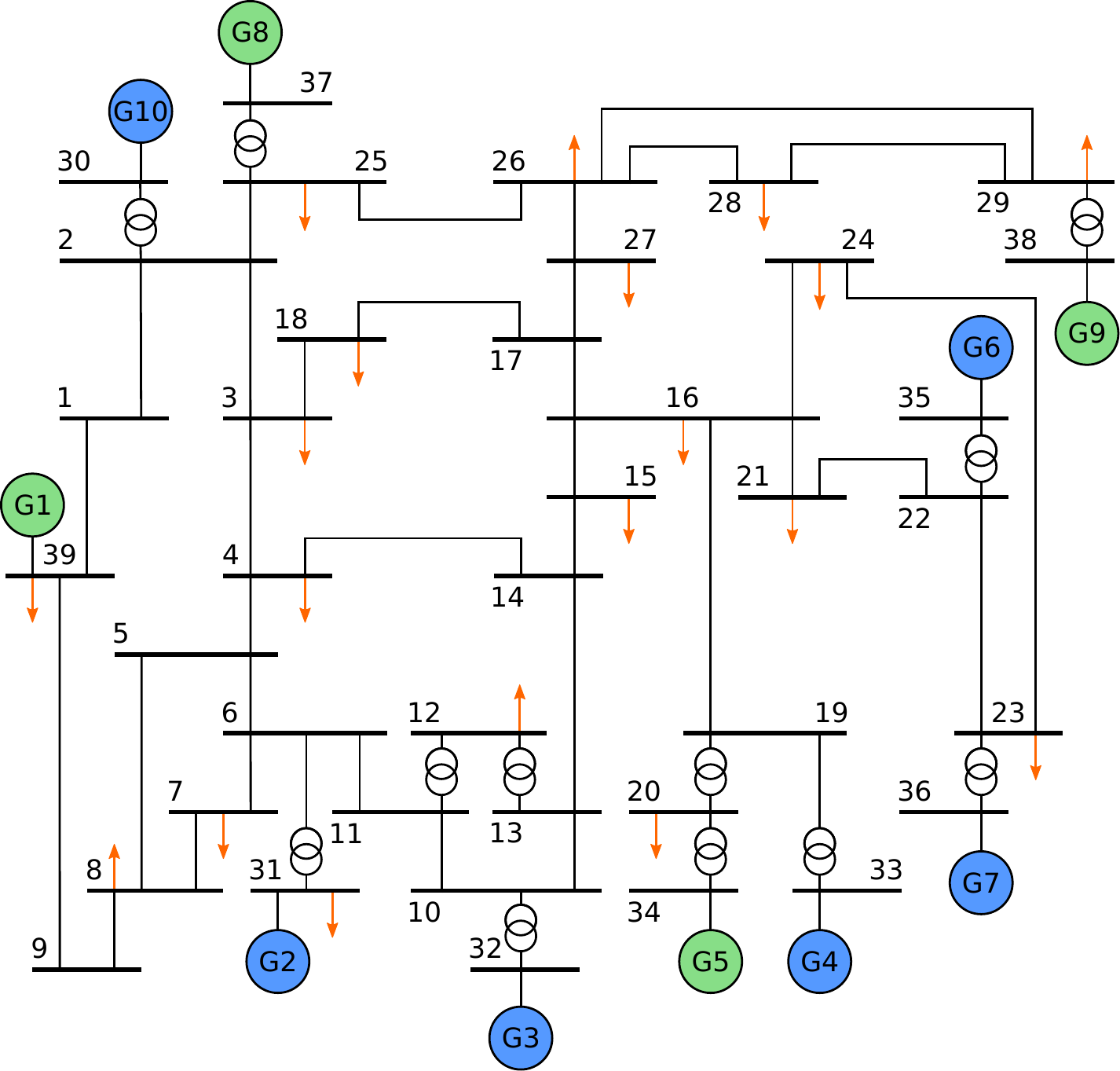}
	\caption{Block diagram of the modified 39-bus power grid Opal-RT simulation model \cite{github39bus}. The system is characterized by nominal voltage of 345 kV. The model consist of 6 synchronous generators (blue circles), 4 wind farms (green circles) and 19 loads (orange arrows). The total installed capacity is 10 GW. }
	\label{fig:IEEE39bus}
\end{figure}

In order to test the proposed technique over large-scale power systems, we adopt the Opal-RT eMEGAsim PowerGrid Real Time Simulator (RTS) \cite{Opal-RT} to implement a detailed dynamic model of IEEE 39-bus power system, also known as 10-machine New-England power system \cite{IEEE39bus}, represented in Fig. \ref{fig:IEEE39bus}. 
This model represents a widely-employed benchmark for performance evaluation and comparison of several monitoring and control applications. In more detail, the simulated power system has a nominal voltage of 345 kV, and consists of 39 buses, 10 generators and 19 loads. 
In order to take into account the effects of distributed renewable generation, we modified the benchmark by replacing 4 conventional synchronous generators with wind farms \cite{Zuo-etAl2018-ISGT}. 
Specifically, this reduced-inertia model includes wind farms instead of generators G1, G5, G8, and G9, for a total wind installed capacity of 4 GW. The total capacity for conventional synchronous generators is 6 GW, for an overall system installed capacity of 10 GW. 
Moreover, in order to emulate realistic load and generation patterns, we used wind and load profiles coming from real measurements. The network is modeled in Simulink and the simulations are run using the Opal-RT eMEGAsim RTS. More details about this model are provided in \cite{github39bus}.

\subsection{The Validation Algorithm}
\begin{figure}
	\centering
	\includegraphics[width=0.87\columnwidth]{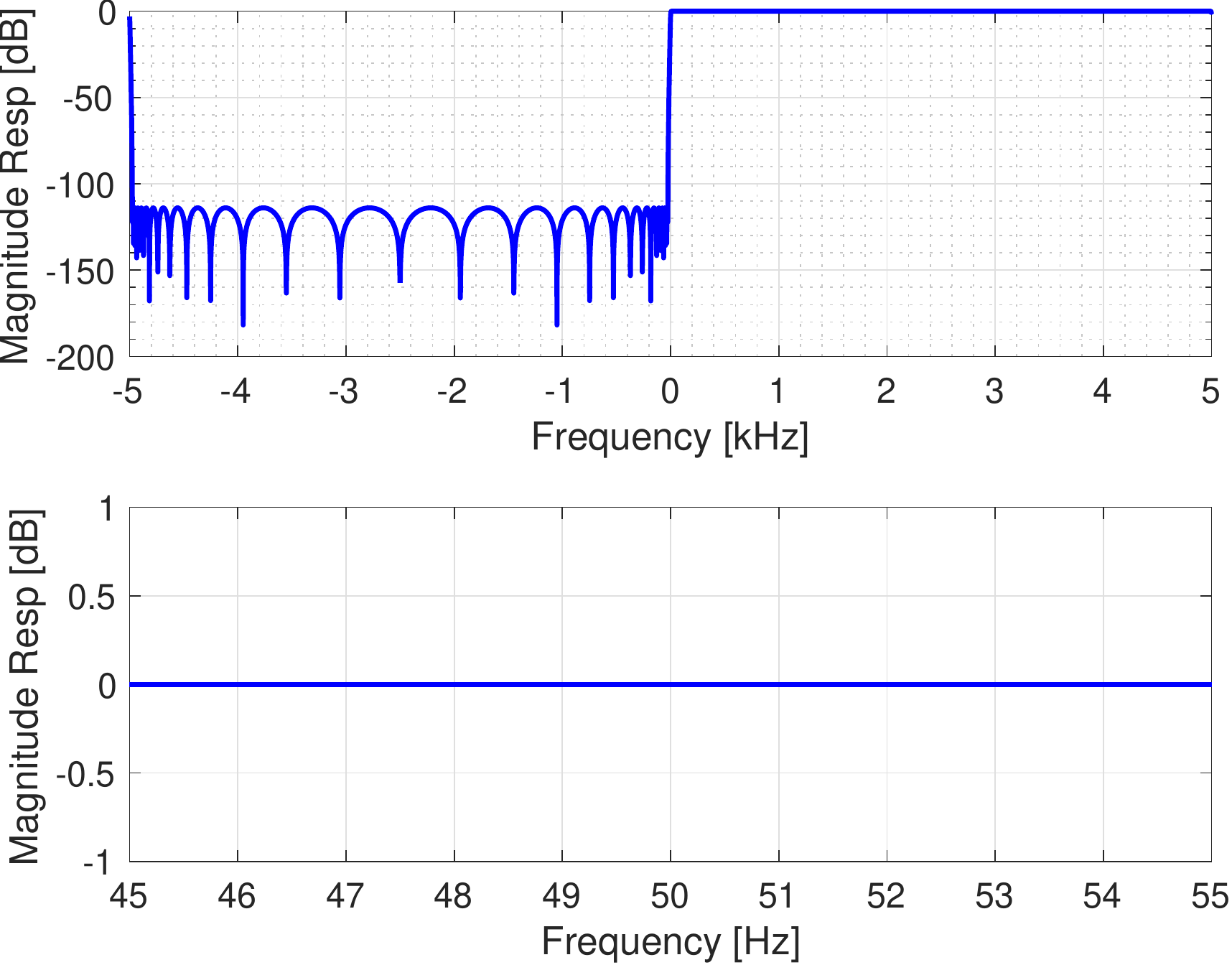}
	\caption{Magnitude response of the filter approximating the ideal HT over the whole frequency bandwidth (a) and over the power system operating conditions (b): filter order 31 and transition width 50$\pi$ radians per sample.}
	\label{fig:filter}
\end{figure}
In the simulated scenario, we know the time-domain 3-phase waveforms of voltage and current in each node, and we post-process them via FT and HT, as described in Algorithm \ref{code:ht}.

\begin{algorithm}
	\caption{}
	\label{code:ht}
	\begin{algorithmic}[1]
		\renewcommand\algorithmicdo{}
		\renewcommand\algorithmicthen{}
		\State $ p(t) = \sum_{abc} v(t) \cdot i(t) $\Comment{True 3-phase  power}  \label{code:p_true}
		\If{$\mathcal{F}[\cdot]$} \Comment{FT computation}
		\For{$ x(t) = \{ v(t), i(t) \}_{abc}$}
		\State $ x(t) = x(t) + \mathcal{N}(t) $ \Comment{Noise adding} \label{code:noise_f}
		\For{$ t=0 \to end, \Delta t = 200 ms$} \label{code:ft}
		\State $ X(k) = \mathcal{F}[x(t) \cdot w(t)]  \approx FFT[x(t) \cdot w(t)]$ \label{code:dft}
		\State $ \{\tilde{f}(t),\tilde{A}(t),\tilde{\varphi}(t)\} = IpDFT(X(k)) $ \label{code:ipdft}
		\State $ \tilde{x}_{\mathcal{F}}(t) = \tilde{A}(t)cos(2 \pi\tilde{f}(t) \Delta t/2 + \tilde{\varphi}(t))  $ \label{code:x_est_f}
		\EndFor
		\State $ \tilde{p}_{\mathcal{F}}(t) =\sum_{abc} \tilde{v}_{\mathcal{F}}(t) \cdot \tilde{i}_{\mathcal{F}}(t)$ \label{code:p_est_f}
		\State $ \Delta p_{\mathcal{F}}(t) = \tilde{p}_{\mathcal{F}}(t) - p(t) $ \label{code:p_err_f}
		\EndFor
		\EndIf
		\If{$\mathcal{H}[\cdot]$} \Comment{HT computation}
		\For{$ x(t) = \{ v(t), i(t) \}_{abc}$}
		\State $ x(t) = x(t) + \mathcal{N}(t) $ \Comment{Noise adding} \label{code:noise_h}
		\State $ \hat{x}_{\mathcal{H}}(t) = x(t) + j\cdot \mathcal{H}[x(t)] \approx filter[x(t)] $ \label{code:x_ana_h}
		\EndFor
		\State $ \hat{p}'(t) = \sum_{abc} \hat{v}(t) \cdot \hat{i}(t) $ \label{code:p1}
		\State $ \hat{p}''(t) = \sum_{abc} \hat{v}(t) * \hat{i}(t) $ \label{code:p2}
		\State $ \hat{p}'''(t) = \hat{p}'(t) + \hat{p}''(t) $ \label{code:p3}
		\State $ \tilde{p}_{\mathcal{H}}(t) = real (\hat{p}'''(t))/2 $ \label{code:p_est_h}
		\State $ \Delta p_{\mathcal{H}}(t) = \tilde{p}_{\mathcal{H}}(t) - p(t) $ \label{code:p_err_h}
		\EndIf
	\end{algorithmic}
\end{algorithm}

First, we define the true instantaneous power $p(t)$ as the product between voltage and current waveforms (line \ref{code:p_true} in Alg. \ref{code:ht}). We focus our investigations on the instantaneous power, to avoid the non-unique interpretation of active and reactive power (and the underlying hypothesis of steady-state phasor) in case of a broad signal spectrum \cite{Ferrero-etAl1991-TIM}. The 3-phase power is obtained by summing the contribution of each phase $\{a,b,c\}$. 
In order to emulate more realistic operating conditions, we simulate the noise and measurement uncertainty by means of a purely additive and uncorrelated white Gaussian noise component (line \ref{code:noise_f} and \ref{code:noise_h}). Specifically, we consider a signal-to-noise ratio (SNR) equal to 80 dB. 

When the FT is adopted, we analyze the signals using a method inspired by the IEC Std. 61000-4-7 \cite{iec61000}: by means of sliding windows of 200 ms, we shift the observation interval sample-by-sample (given the sampling frequency of 10 kHz, this corresponds to steps of 100 $\mu$s) (line \ref{code:ft}). 
For each window, we approximate the FT by means of the Discrete FT (DFT) computed via a Fast FT (FFT) algorithm (line \ref{code:dft}). 
The adopted window length (corresponding to 10 cycles at the nominal power system frequency) determine a 5 Hz granularity in the frequency domain. 
In order to reduce  spectral leakage effects, we window the signal with a Hanning weighing function $w(t)$. 
In order to reduce the effects of spectrum granularity, for each observation window we define the estimated frequency $\tilde{f}(t)$, amplitude $\tilde{A}(t)$ and phase $\tilde{\varphi}(t)$ by means of the Interpolated DFT (IpDFT) (line \ref{code:ipdft}) \cite{Jain-etAl1979-TIM,Grandke-etAl1983-TIM}.
For the sake of conciseness, this analysis is not repeated for different window lengths or windowing functions\footnote{It is worth observing that as for any FT-based approach, the longer the observation window, the smaller the spectrum granularity and thus the higher the frequency domain accuracy. By contrast, a shorter window length results in a faster response time in passing from one steady state to the other. In this study we are mainly interested in the accuracy of the estimates, rather than on their responsiveness, therefore, we adopt a fairly long observation interval.}. 
These estimates are used to approximate the time domain trend of the sinusoidal component in the considered observation interval. 
We construct the estimated signal $\tilde{x}_{\mathcal{F}}(t)$ by extracting the central point of each of the consecutive sliding windows (line \ref{code:x_est_f}). 
The 3-phase power is obtained by summing the contribution from each phase (line \ref{code:p_est_f}) resulting in power errors (line \ref{code:p_err_f}).

As regards the HT, we first compute the analytic signal of the noisy waveforms (line \ref{code:x_ana_h}). 
We approximate the analytic signal by means of a filter using 10 kHz sampling frequency, filer order set to 31 and transition width set to 50$\pi$ radians per sample. As shown in Fig. \ref{fig:filter}, the magnitude response is nearly equal to zero (the ideal value) for the positive frequency domain, whereas it is lower than -120 dB in the negative frequency range, achieving a high rejection of the long-range interference coming from the negative frequency components. 

An alternative solution for computing the signal HT is provided by functional analysis. 
The acquired signal $x(t)$ is projected over a vector basis whose kernel is defined as:
\begin{equation}
    (1 + g_A(t))\cdot \cos(2\pi g_f(t) t + g_\phi(t))
    \label{eq:basisfunc}
\end{equation}
where the functions $g_A$, $g_f$ and $g_\phi$ account for the time-variations of amplitude, frequency and phase of the fundamental component respectively, and might follow different trends like ramps, sinusoids or step changes. 
Differently from the filtering approach, the basis projection would allow to identify a mathematical model of the parameters' evolution that, de facto, enables compressing time domain information into few coefficients of \eqref{eq:basisfunc}. However, its performance depends on the number of possible realizations of \eqref{eq:basisfunc} included in the basis: the larger the more accurate, but also the more computationally demanding and numerically ill-conditioned. In this paper we decided to focus on power estimation via the filter in Fig. \ref{fig:filter}.

Finally, we compute the quantities defined in equations \eqref{eq:p1}-\eqref{eq:p3} (line~\ref{code:p1}-\ref{code:p3}). 
As in \eqref{eq:pht}, the estimated power $\tilde{p}_{\mathcal{H}}(t)$ is computed as the real part of $\hat{p}'''(t)$ (line~\ref{code:p_est_h}) and the corresponding power errors are consequently computed (line~\ref{code:p_err_h}).

\section{Results}
\label{sec:results}


In this section, we describe the results that enable us to validate the proposed HT-based analysis. 
We analyze three different datasets. 
The first one is obtained using EMTP-RV and is inspired by the signals formulated in Section \ref{sec:theory}, representing approximations of actual power system operating conditions. 
The second dataset is still obtained using EMTP-RV and refers to real-world events. 
Particularly, we replicate the waveforms taking place in Australia in September 2016 (see Fig.~\ref{fig:intro}) and in Europe during an inter-area oscillation in December 2016 \cite{SIL}. In that occasion, an unexpected opening of a line in the French transmission network caused a voltage phase angle difference in the continental Europe electricity system and decreased the general damping that triggered a permanent oscillation at 0.15 Hz \cite{SIL}.
The third set of waveforms is obtained simulating a large contingency within the 39-bus model using Opal-RT. 

For each operating condition, the results are presented by means of two plots showing the instantaneous active power computed using the FT and the HT as in Alg. \ref{code:ht}. The upper plot represents the power in phase $a$ and the true reference power, the bottom plot represents the 3-phase power error. The results are presented for one bus only, but similar considerations hold for the other buses of the considered network. 

\subsection{Theoretical Operating Conditions}

By using Matlab environment, we numerically simulate a plausible situational awareness context, with a sampling frequency of 10 kHz. We synthesize voltage waveforms with a nominal voltage of 380 kV, for a total duration of 4 seconds for each test. 
These waveforms are used to model the output of the voltage source are presented in Fig.~\ref{fig:model}. 
Then, as discussed in Section~\ref{sec:model}, we analyze voltage and current signals as provided by EMTP-RV simulations. 
In more detail, we consider three cases with the signal characterized by:
\begin{enumerate}[a)]
	\item Amplitude modulation, being $f_a$ = 5 Hz the modulation frequency and $k_a$ = 0.1 the modulation depth in~\eqref{eq:am};  
	\item Negative frequency ramp in the range $50 \leq f_0 \leq 46$, being $R$ = $-$6 Hz/s the ramp rate in~\eqref{eq:fr};
	\item Amplitude step, being $k_s$ = 0.1 the step factor in~\eqref{eq:as}. 
\end{enumerate}

\begin{figure}
	\centering
	\includegraphics[width=0.87\columnwidth]{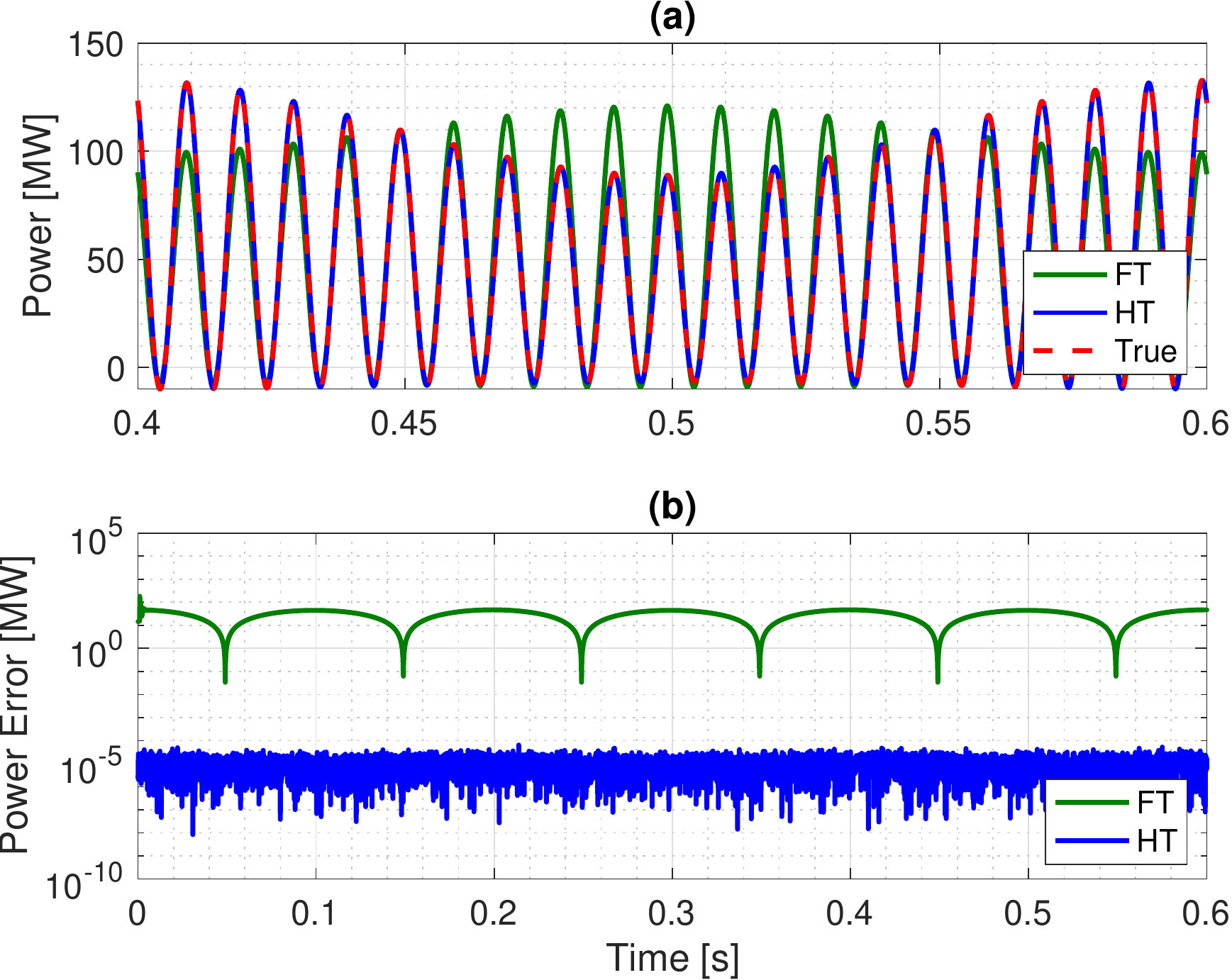}
	\caption{Amplitude modulation: instantaneous single-phase active power (a) and three-phase power error (b) computed using the FT (green) and the HT (blue) in case of a signal characterized by an amplitude modulation with modulating frequency of 5 Hz. The red line represents the true power.}   
	\label{fig:p_am}
\end{figure}

Regarding the amplitude modulation case, Fig.~\ref{fig:p_am} shows that the FT does not enable us to correctly interpret a signal whose fundamental component is modulated. 
\begin{figure}
	\centering
	\includegraphics[width=0.87\columnwidth]{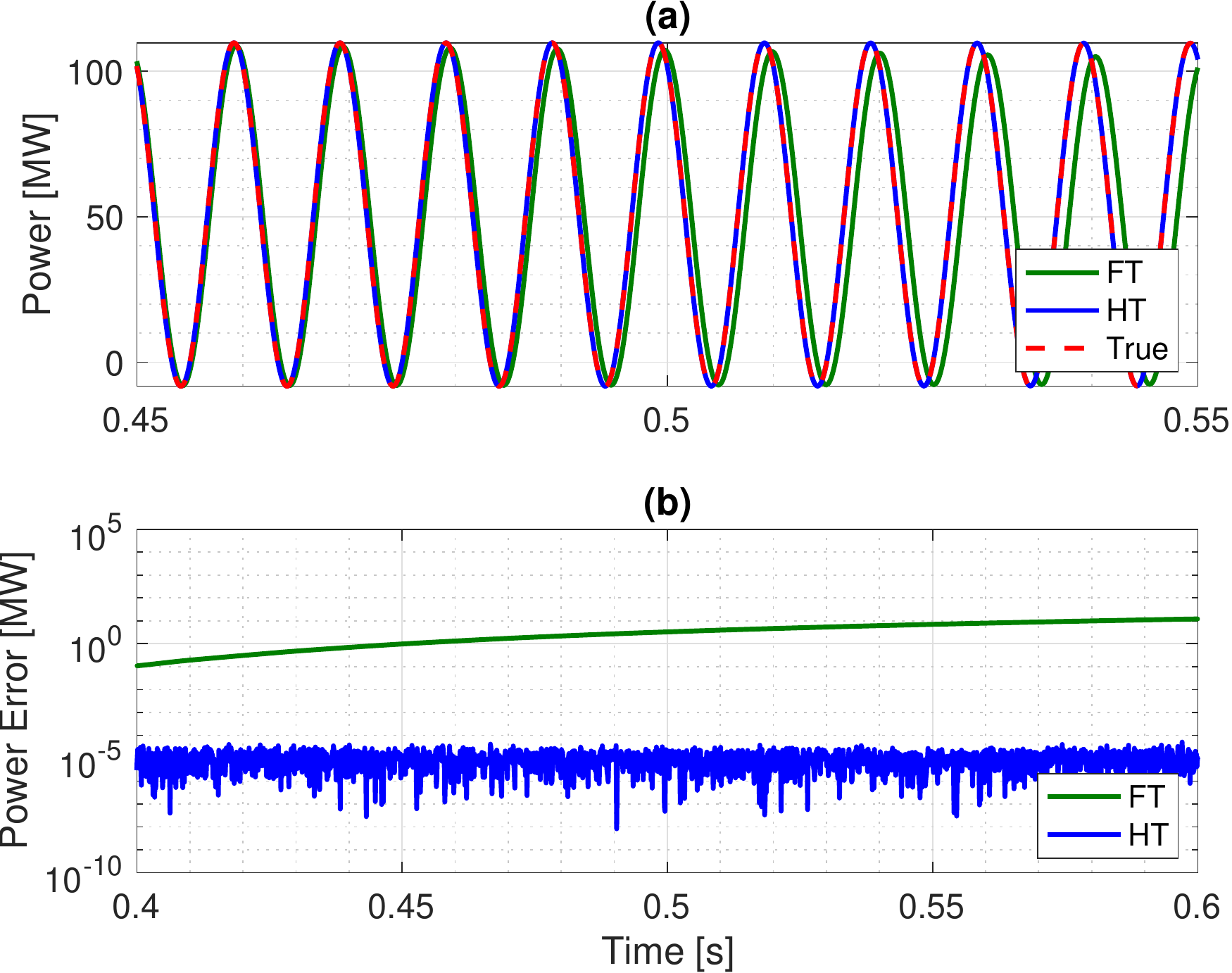}
	\caption{Frequency Ramp: instantaneous single-phase active power (a) and three-phase power error (b) computed using the FT (green) and the HT (blue) in case of a signal characterized by a negative frequency ramp of -6 Hz/s. The red line represents the true power.} 
	\label{fig:p_fr}
\end{figure}
\begin{figure}
	\centering
	\includegraphics[width=0.87\columnwidth]{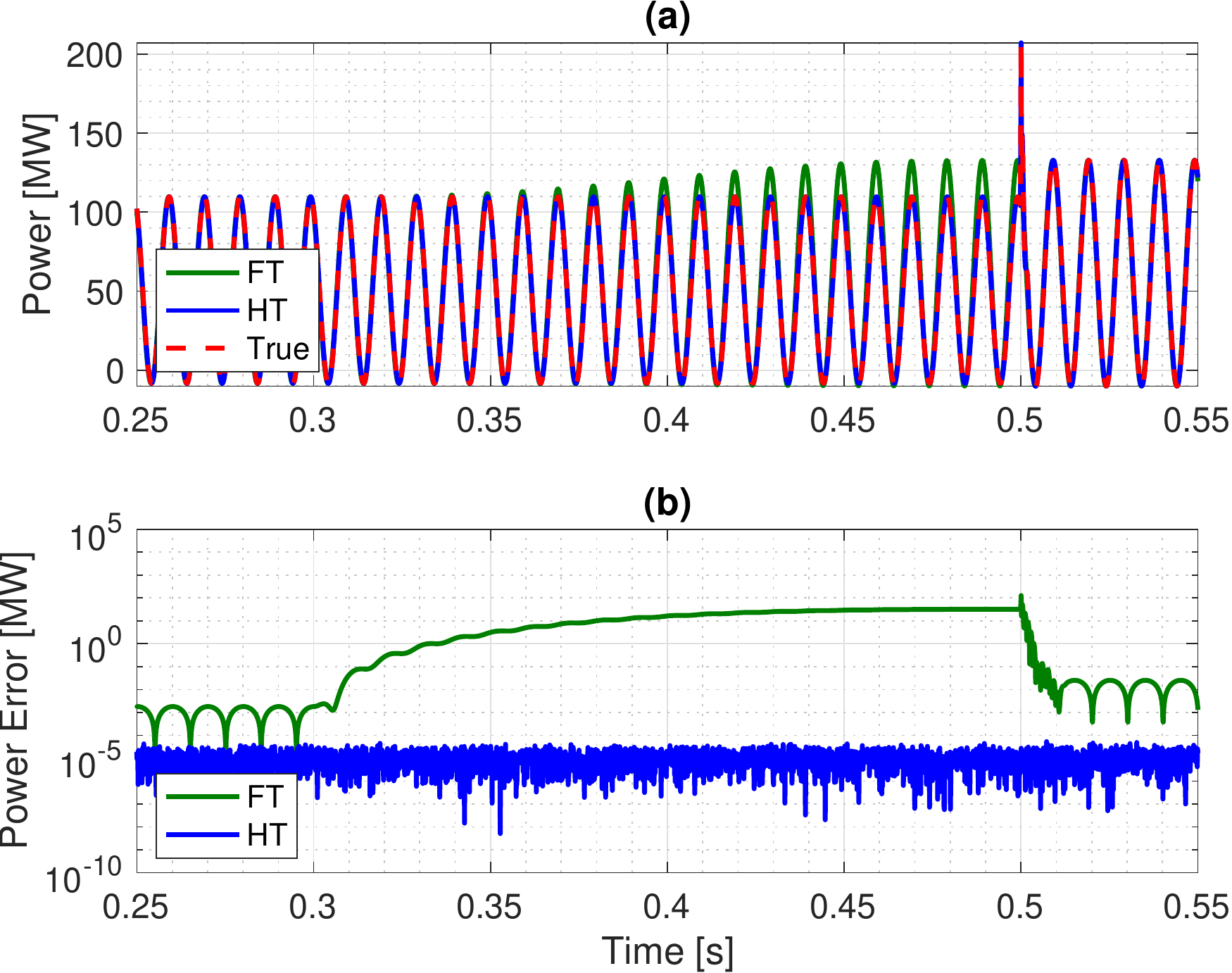}
	\caption{Amplitude Step: instantaneous single-phase active power (a) and three-phase power error (b) computed using the FT (green) and the HT (blue) in case of a signal characterized by a 10\% amplitude step. The red line represents the true power.} 
	\label{fig:p_as}
\end{figure}
Indeed, the obtained signal spectrum is largely biased by the interference produced by the modulating term, leading to imprecise parameters estimation. Both voltage and current are inaccurately measured, leading to a maximum power error in the order of 50 MW. 
By contrast, it is evident that the HT preserves the 
information needed to entirely reconstruct the analyzed signal: the maximum attained power errors are in the order of 10 W.

Regarding the frequency ramp case, Fig. \ref{fig:p_fr} shows that as the frequency deviates from its nominal value, the FT provides erroneous results due to spectral leakage.
The maximum obtained power error is in the order of 10 MW. Conversely, the HT is able to follow instantaneously the signal dynamics experienced during a frequency ramp, providing an almost-perfect signal reconstruction. Also in this operating condition, the HT provides errors lower than 10 W.

Regarding the amplitude step case, the FT spectrum does not contain the information needed to infer the signal parameters during such sudden waveform deformations. 
Fig. \ref{fig:p_as}, shows that over the whole period during which the step is contained in the sliding window (i.e., 200 ms), the FT provides wrong results reaching 30 MW error when the step appears in the center of the window and therefore the spectrum reconstruction is the most wanting. On the contrary, the HT enables us to follow the signal throughout the whole time interval, matching almost sample-by-sample the signal parameters. Again, the maximum attained power error is in the order of 10 W. 

\subsection{Real-world Operating Conditions}
The second set of waveforms, refers to  real-world events. 
Particularly, we replicate the waveforms taking place in Australia during the blackout in September 2016 and in Europe during the inter-area oscillations in December 2016 \cite{AEMO,SIL}. Based on PMU estimates of fundamental frequency, amplitude and initial phase, we recover the time-domain waveform as sampled at 10 kHz, through the approach in \cite{Frigo-etAl2017}. These datasets are used to replicate the voltage source in Fig. \ref{fig:model}. 

\begin{figure}
	\centering
	\includegraphics[width=0.87\columnwidth]{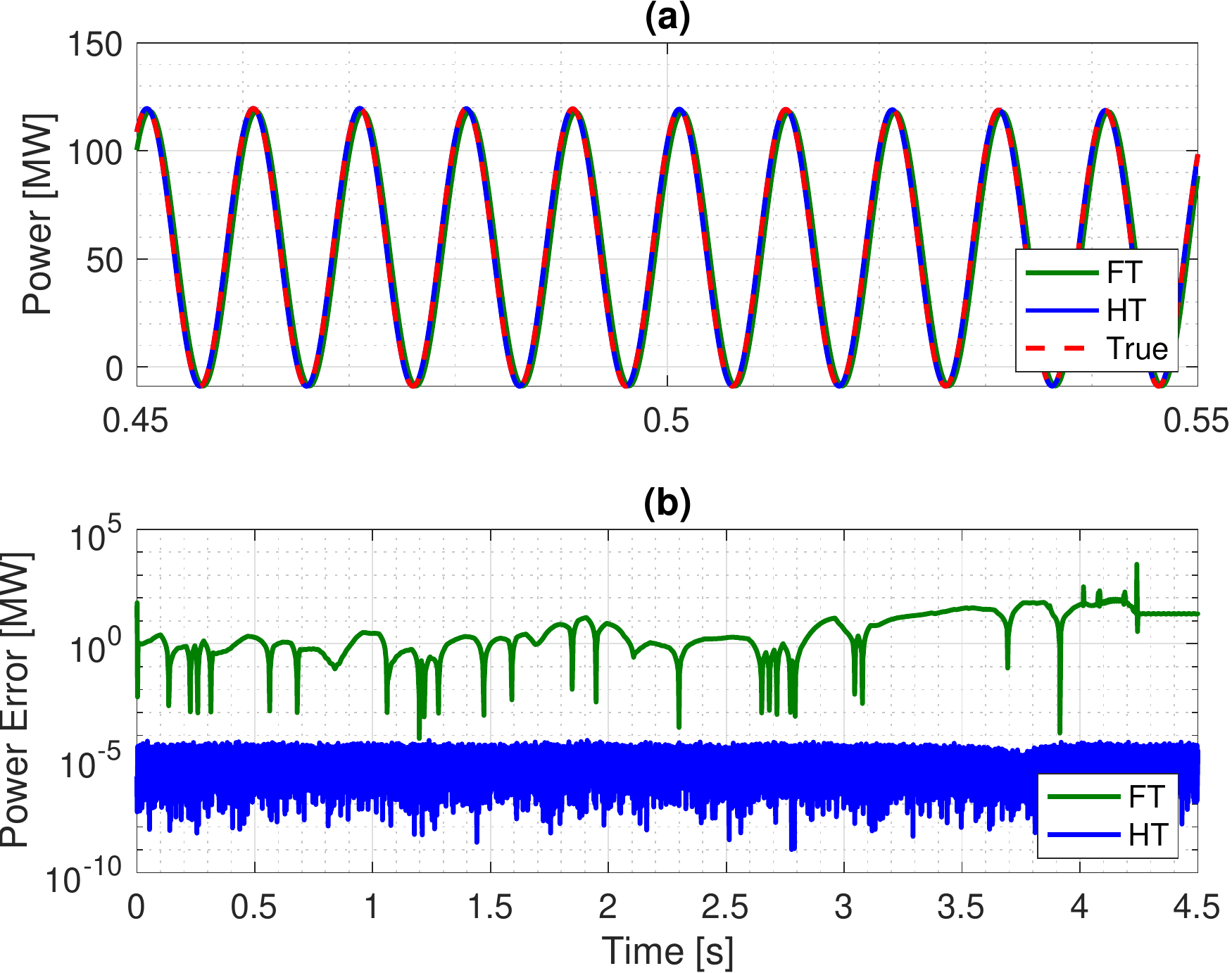}
	\caption{Australian blackout: instantaneous single-phase active power (a) and three-phase power error (b) computed using the FT (green) and the HT (blue) in case of replicating the waveforms that took place in Australia during the blackout on September 28, 2016. The red line represents the true power.}
	\label{fig:p_aus}
\end{figure}

Regarding the Australian blackout, as illustrated in Fig \ref{fig:intro}, the signal is characterized by two criticalities: first the amplitude step occurring at 1.6 s, then the sudden frequency drop at 2.7 s. As discussed so far, the two events are difficult to analyze adopting FT-based tools because the signal spectrum is so spread that it is ambitious to infer the waveform parameters. Conversely, the HT has proven to be potentially suitable for analyzing both situations. As a matter of fact, the results in Fig. \ref{fig:p_aus} confirm the inappropriateness of the FT to represent both the amplitude step and the frequency drop, exhibiting errors in the order of 100 MW. Besides, the HT is characterized by a maximum error lower than 10 W.

\begin{figure}
	\centering
	\includegraphics[width=0.87\columnwidth]{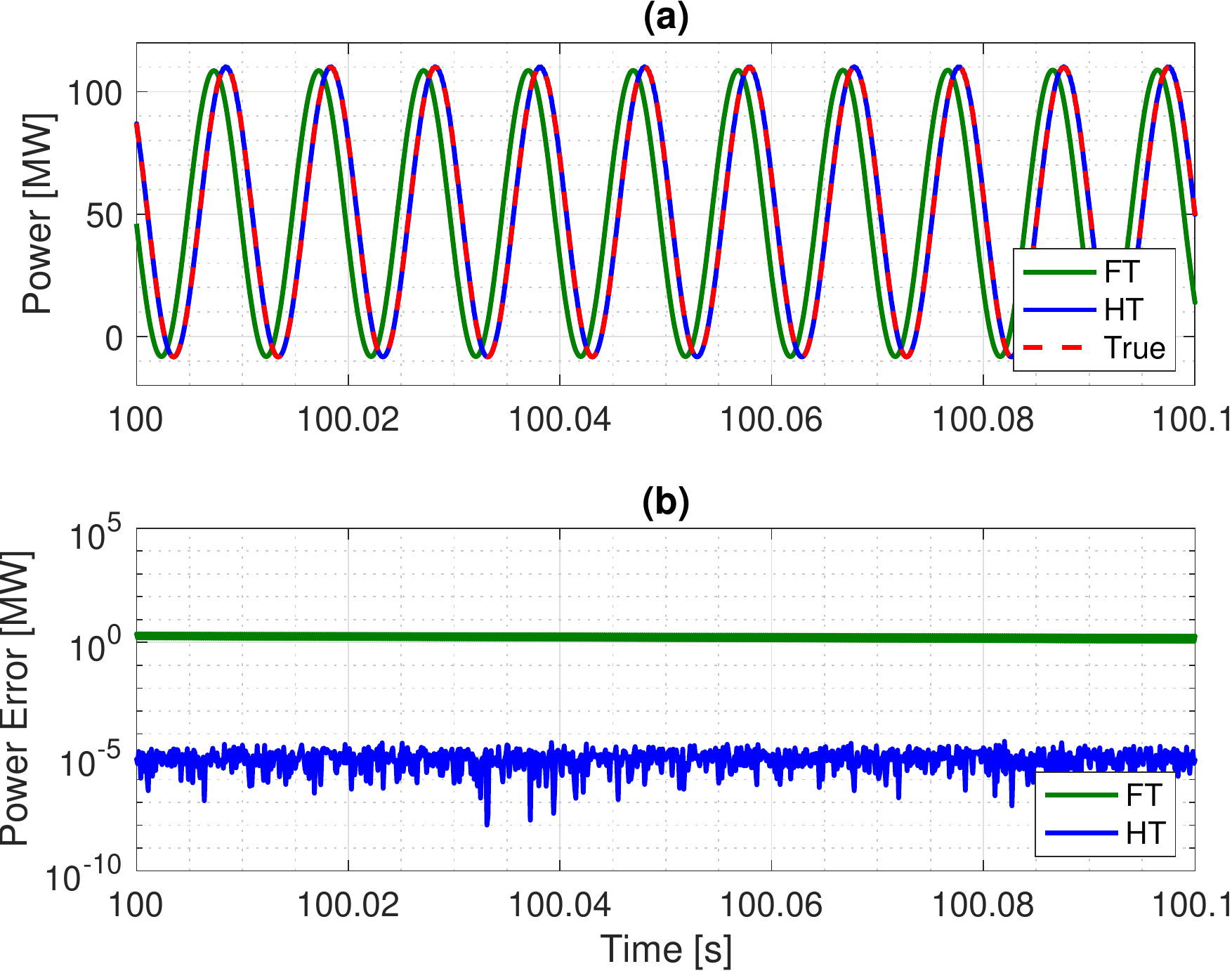}
	\caption{European inter-area oscillation: instantaneous single-phase active power (a) and three-phase power error (b) computed using the FT (green) and the HT (blue) in case of replicating the waveforms that took place in Switzerland during the inter-area oscillation on December 1, 2016. The red line represents the true power.}
	\label{fig:p_sil}
\end{figure}

Regarding the European inter-area oscillation, Fig. \ref{fig:p_sil} confirms the improvement of adopting the HT in stead of the FT, as the instantaneous power estimates are always characterized by a smaller error: maximum error lower than 10 W for the HT and 5 MW for the FT. The HT proves to be a suitable tool also for analyzing real-world waveforms. 
\begin{figure}
	\centering
	\includegraphics[width=0.87\columnwidth]{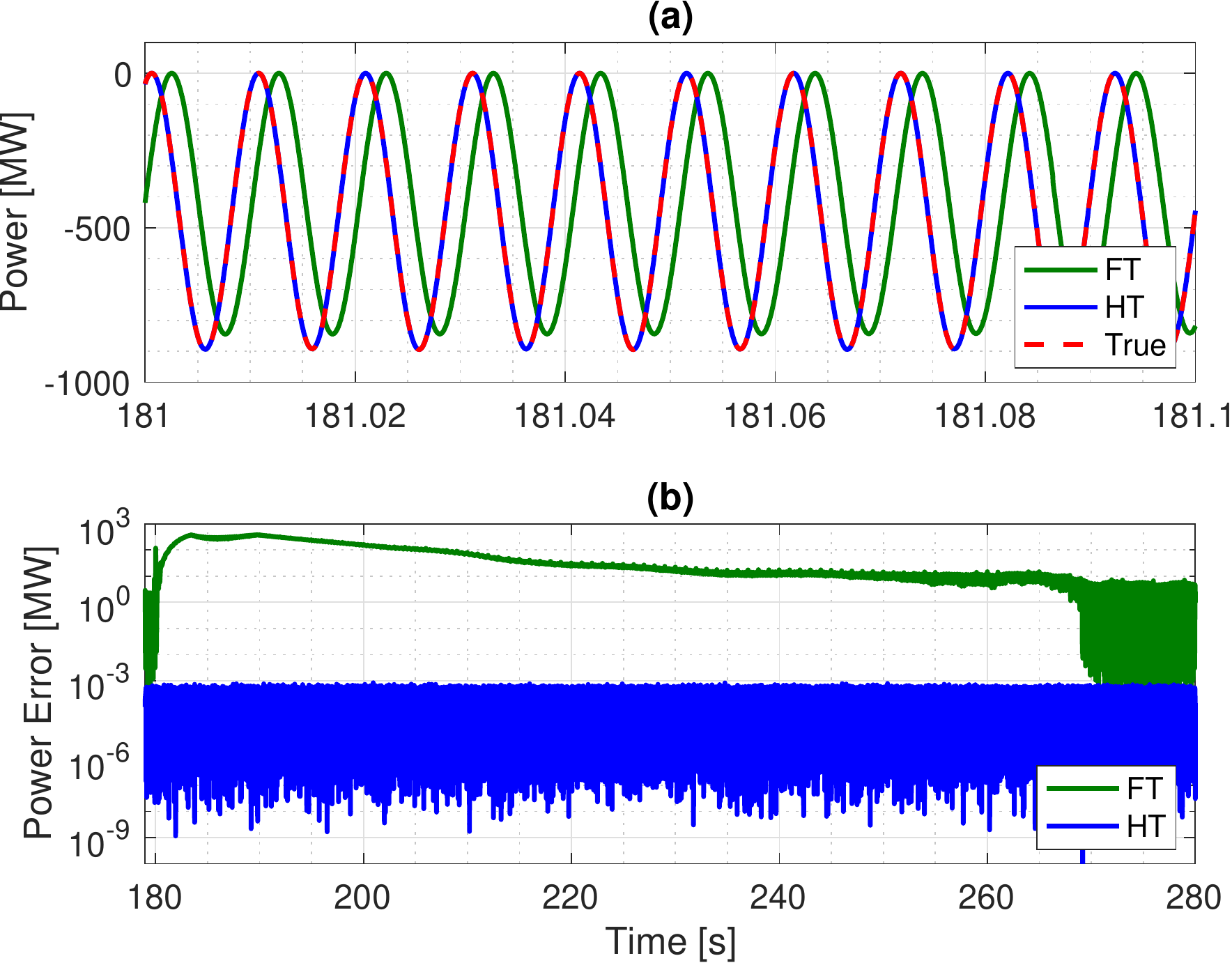}
	\caption{IEEE 39-bus: instantaneous single-phase active power (a) and three-phase power error (b) computed using the FT (green) and the HT (blue) in case of tripping generator G6 at second 180. The power profiles refer to bus 21. The red line represents the true power.}
	\label{fig:p_39bus}
\end{figure}

\subsection{Large-scale Power System}

In order to evaluate the appropriateness of the proposed technique to model large-scale power grids, we carry out dedicated simulations of emergency scenarios using the 39-bus model in Opal-RT environment. 
In particular, we simulate the outage of generator G6, with a total tripped power of 
800 MW, leading a load imbalance that determines a strong system dynamic. The generator is tripped at second 180 and the transient lasts for roughly 100 seconds. 

Fig. \ref{fig:p_39bus} shows the power profiles recorded in bus 21, but similar results hold for all the buses of the network. As it is shown in the figure, during the whole transient the FT does not provide a truthful representation of the power system behavior, leading to errors in the order of 300 MW. Conversely, the HT provides an accurate transient tracking providing errors always lower than 800 W.



\section{Conclusions}
\label{sec:concl}
In this paper, we proposed an HT-based approach for studying broadband power system dynamics as an alternative to the traditional phasor-based  representation. First, we compared the FT and HT of three large power system dynamics, and we discussed the limitations of FT-based analysis in non-stationary conditions, as well as the HT capability of tracking the evolution of the signals. Then, we carried out a numerical analysis where we developed two algorithms for the estimation of instantaneous power, relying on FT and HT representation, respectively. 
We characterized the accuracy to correctly compute the transmitted instantaneous power of the proposed transforms in both synthetic and real-world datasets. 
To this end, we carried out dedicated time-domain simulations of a simple 2-bus system and of the IEEE 39-bus. 

Both theoretical and numerical results confirm that the HT is a suitable tool to be used in power system modeling and operation. 
Indeed, the HT provides superior results when computing the instantaneous power in all the considered operating conditions, leading to power errors up to 100 times smaller than in the case of using an FT-based representation.

The treatise presented in this paper opens new scenarios for modern power systems modeling.
Indeed, the conservation of energy in the HT domain demonstrates the possibility to leverage on HT-based power system analysis, provided that an appropriate functional basis is formulated. 
On the one hand, HT-based situational awareness systems that rely on a broad spectrum could be deployed. For instance, we could think of PMUs capable of computing the analytic signal rather than the fundamental component synchrophasor. 
On the other hand, circuit theory fundamental laws could be formulated in the HT domain. Potentially, we could think of HT-based tools for power flow analysis. 

\bibliographystyle{IEEEtran}
\bibliography{mybibfile}

\end{document}